\documentclass[sigconf, nonacm]{acmart}

\usepackage{subcaption}  
\usepackage{graphicx}
\usepackage{xcolor}
\usepackage{multirow}
\usepackage{balance}

\AtBeginDocument{%
  \providecommand\BibTeX{{%
    \normalfont B\kern-0.5em{\scshape i\kern-0.25em b}\kern-0.8em\TeX}}}

\copyrightyear{2025}
\acmYear{2025}
\setcopyright{cc}
\setcctype{by}
\acmConference[UAVM '25]{Proceedings of the 3rd International Workshop on UAVs in Multimedia: Capturing the World from a New Perspective}{October 27--28, 2025}{Dublin, Ireland}
\acmBooktitle{Proceedings of the 3rd International Workshop on UAVs in Multimedia: Capturing the World from a New Perspective (UAVM '25), October 27--28, 2025, Dublin, Ireland}\acmDOI{10.1145/3728482.3757388}
\acmISBN{979-8-4007-1839-7/2025/10}

\begin{document}

\title{Illuminating English Letters Using a Flying Light Speck}




\author{Hamed Alimohammadzadeh}
\email{halimoha@usc.edu}
\orcid{0000-0003-2613-5010}
\affiliation{%
  \institution{University of Southern California}
  \city{Los Angeles}
  \state{CA}
  \country{USA}
}

\author{Shahram Ghandeharizadeh}
\email{shahram@usc.edu}
\orcid{0000-0002-1792-7879}
\affiliation{%
  \institution{University of Southern California}
  \city{Los Angeles}
  \state{CA}
  \country{USA}
}

\begin{abstract}
This paper presents the design and implementation of a Flying Light Speck (FLS) to illuminate English letters.
The FLS uses its onboard camera and computing to localize and follow a trajectory to illuminate a letter.
We evaluate the illuminations quantitatively and qualitatively.
The latter is based on an IRB approved human subject study with 20 participants.
The obtained results show a 42 to 56 millimeter error that impacts the detection of letters.
A key finding is that the order in which the illumination of letters is presented to subjects has a significant effect on detection duration.

\end{abstract}

\ccsdesc[500]{Human-centered computing~Visualization design and evaluation methods}
\ccsdesc[500]{Human-centered computing~User studies}
\ccsdesc[500]{Computing methodologies~Tracking}
\ccsdesc[500]{Computing methodologies~Motion path planning}
\ccsdesc[500]{Computer systems organization~Embedded and cyber-physical systems}
\ccsdesc[500]{Information systems~Multimedia content creation}

\keywords{Flying Light Speck, Illumination, English Letters}



\begin{teaserfigure}
    \centering
    \hfill
    \begin{subfigure}[t]{0.25\textwidth}
        \centering
        \includegraphics[width=0.5\textwidth]{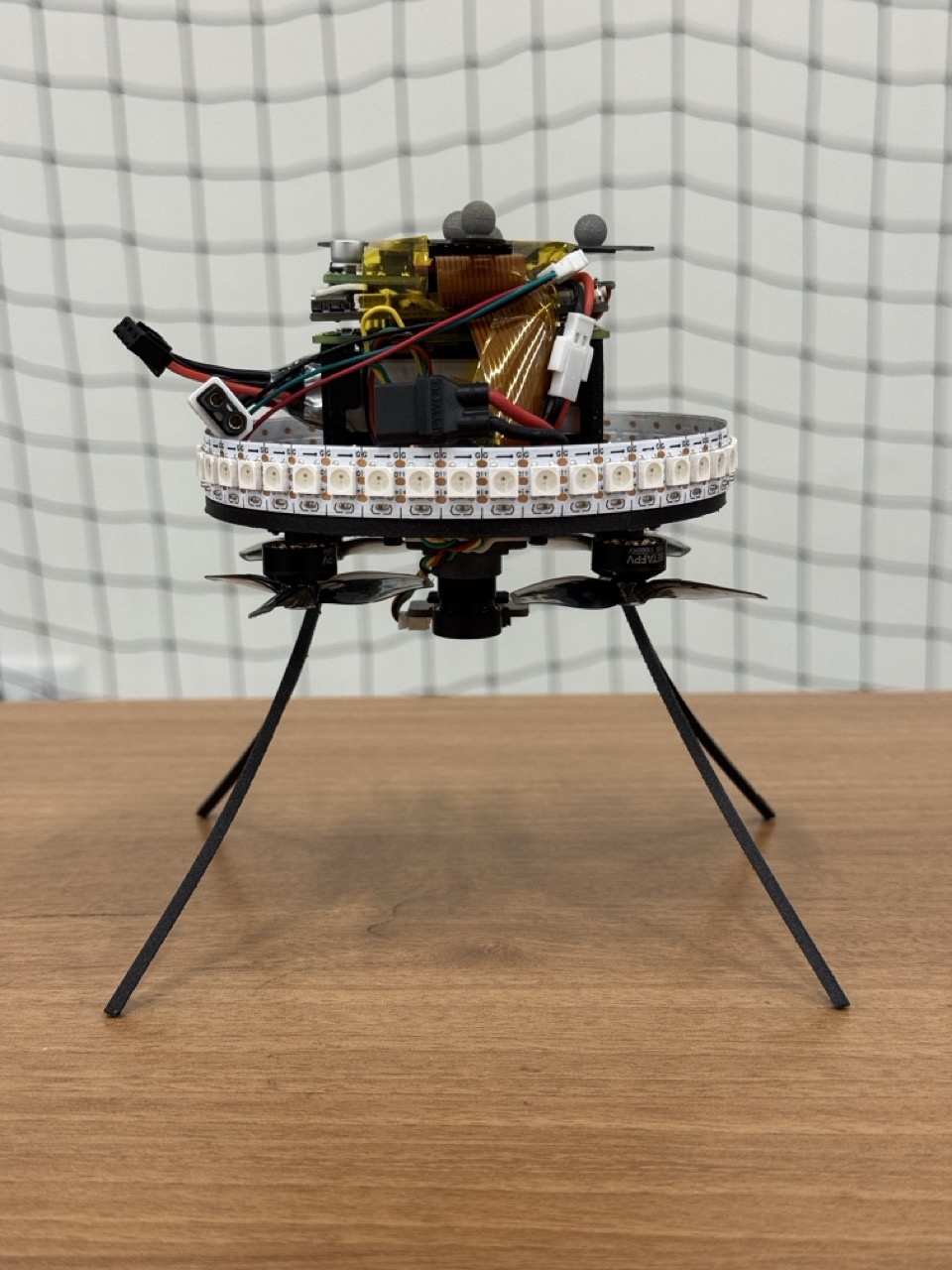}
        ~\includegraphics[width=0.5\textwidth]{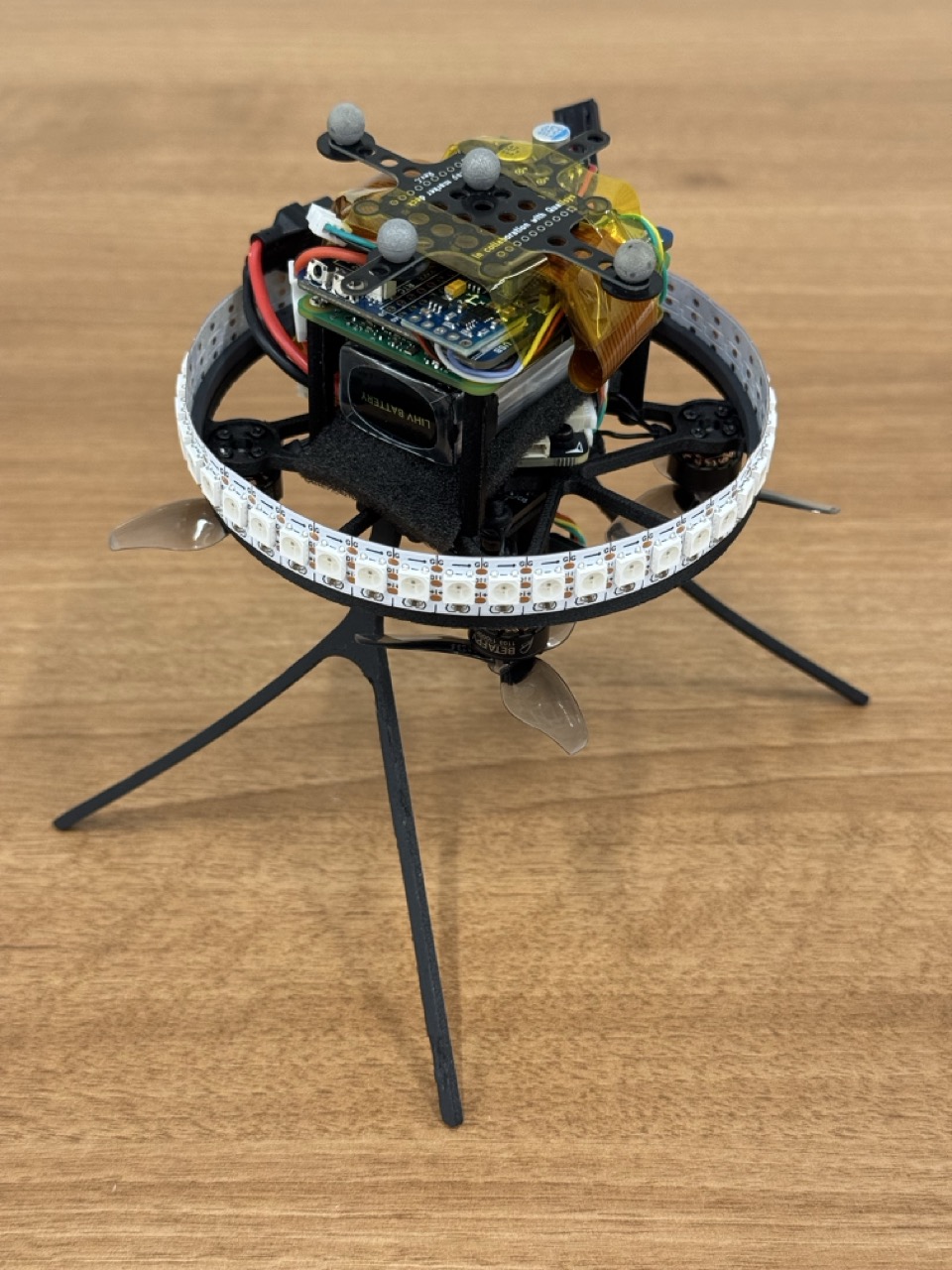}
        \caption{ }
        \label{fig:fls}
    \end{subfigure}
    \hfill
    \begin{subfigure}[t]{0.25\textwidth}
        \centering
        \includegraphics[width=\textwidth]{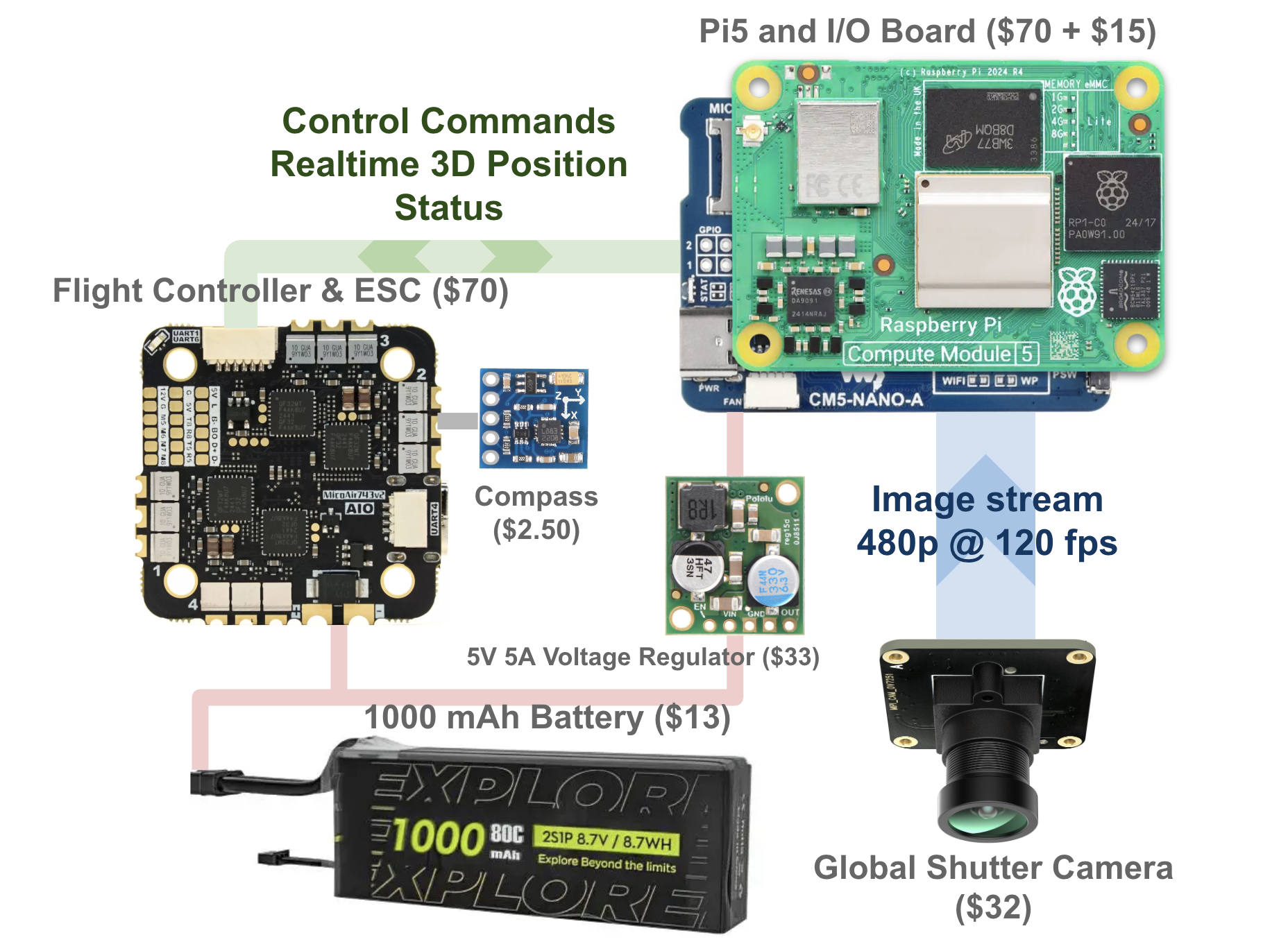}
        \caption{ }
        \label{fig:diagram}
    \end{subfigure}
    \hfill
    \begin{subfigure}[t]{0.45\textwidth}
        \includegraphics[width=\textwidth]{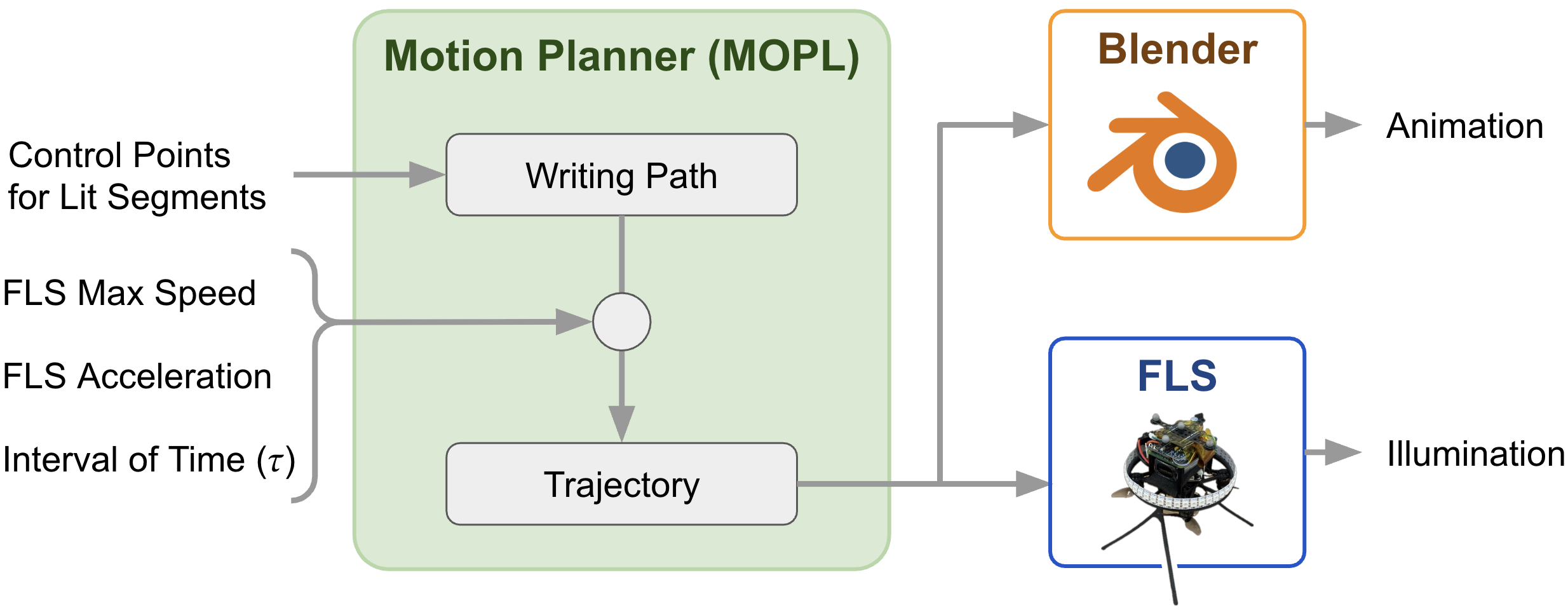}
        \caption{ }
        \label{fig:overall}
    \end{subfigure}
  \caption{(a) The FLS prototype with LED ring. (b) Diagram of FLS components. (c) Process of animating and illuminating letters.}
\end{teaserfigure}


\maketitle

\section{Introduction}

A Flying Light Speck (FLS) is a small drone equipped with one or more light sources that can generate different colors and textures with adjustable brightness~\cite{shahram2021}.
Swarms of FLSs illuminate complex 2D and 3D shapes within a fixed volume, essentially creating a 3D display~\cite{shahram2022, decentralized2023, mmsys2023, dv2023,flightpatterns2023,flshaptic23, flshaptics2023, flshaptics2024, circular2024, reliability2024, tomm2025,integrate2023,integrate2025,mcgeKeynote2025}.
FLSs position themselves relative to one another by communicating~\cite{swazure2024} to implement a decentralized localization technique~\cite{swarmer2023,swarical2024,swaricalrepo2025}.

In this workshop~\cite{wang2025UVA} paper, we present the design and implementation of an FLS, see Figure~\ref{fig:fls}.
We use this FLS to illuminate English letters such as E and N.
The process of creating these illuminations is shown in Figure~\ref{fig:overall} and is as follows. 
A human designer authors control points and segments that define the writing path for a letter.
These are processed by our {\em Motion Planner} software (MOPL) to compute a trajectory for the FLS.
MOPL uses the maximum FLS speed, its acceleration, and a time interval $\tau$ to create the trajectories.
We use Blender to animate the trajectories. 
This animation is a form of visualization to evaluate a trajectory and to refine it prior to illuminating it using an FLS.
An FLS uses its decentralized localization technique to fly along a trajectory.




Illuminating a letter using an FLS is non-trivial.  A letter such as N consists of three lines that join at one end, requiring the FLS to make a sharp turn.
A different letter such as S requires the FLS to travel along a curve.
The start and end points of both letters are at different coordinates.
The FLS must turn its light on while navigating from start to end to illuminate the letter and turn its light off when flying from end to start to repeat the illumination.
An FLS must do all this without the use of GPS as it is indoors.
And, with minimal deviation from its specified trajectory.
Otherwise, a human subject will not be able to detect an illumination as a letter.

{\bf Contributions} of this paper include:
\begin{enumerate}
\item Design and implementation of an FLS with a relative decentralized localization technique.
Section~\ref{sec:fls}.
\item Illumination of 4 English letters using one FLS. Section~\ref{sec:letter}.
\item A quantitative and qualitative analysis of the FLS illuminated letters.  
Section~\ref{sec:eval}.
\item We open source the design of our FLS and its software at \url{https://github.com/flyinglightspeck/FLS}.
\end{enumerate}
We present related work in Section~\ref{sec:related} and future work in Section~\ref{sec:future}.

\section{FLS Design and Implementation}\label{sec:fls}

An FLS is a miniature drone designed to perform localization using only its onboard sensors \cite{imeta2023}. Our design focuses on a decentralized localization technique \cite{swarical2024}, implemented using a global shutter camera and a compute module. 
This section describes the FLS hardware, control system, and its vision-based localization technique.

Designed for autonomous operation, the FLS integrates all necessary components while keeping weight and size to a minimum.
Its hardware components are shown in Figure~\ref{fig:diagram}.
The frame and body parts are 3D printed using a light and durable PA6-CF filament. The final prototype depicted in Figure \ref{fig:fls} weighs 153 grams with a motor-to-motor distance of 6 cm.

The Pi5 is the main computer, responsible for high-level control, localization computations, and system monitoring.
It communicates with the FC through a serial port at a baud rate of 115200. The global shutter camera is also connected to the Pi5. The Pi5 runs two programs concurrently:
First, the marker localization program that captures camera images to calculate the FLS's position. 
Second, the offboard controller program that controls the FLS, issuing commands for takeoff, trajectory following, and landing. It receives the position estimates from the marker localization program and forwards them to the FC at a rate of 120 Hz. It also monitors the battery for safety and controls the LED lights.


To determine its position, the FLS captures images of a custom marker placed on the floor.
The marker consists of four infrared (IR 850 nm) LEDs. The LEDs are arranged in a pattern known to the FLS. An 850 nm narrow bandpass filter is mounted on the camera lens to ensure only the light from the IR LEDs is captured. For each captured image, the system detects the four marker points.
These four points are then used to calculate the 3D position of the FLS relative to the marker.
This is known as the Perspective-n-Point (PnP) Pose Computation problem \cite{pnp}. We use the AP3P technique \cite{ap3p}, implemented in the OpenCV library \cite{opencv_library}, to solve this problem.


\section{English Letters and Segments}\label{sec:letter}

Illuminating a letter using an FLS is a two-stage process, see Figure~\ref{fig:overall}. First, a human designer creates a 
{\em Writing Path}. 
Second, our {\em Motion Planner} software (MOPL) converts this into an executable {\em Trajectory} for the FLS to follow.
The writing path is the blueprint that defines the shape of a letter.
This path is composed of one or more sequential segments. A segment is
a straight line or a curve. Each segment is designated as either {\em lit} or {\em dark}.

As an analogy, consider the movement of a pen when writing a letter on a piece of paper.  The writing path is the movement of the pen on the paper. 
A lit segment is analogous to a pen stroke made on the paper. And a dark segment is analogous to lifting the pen to move to a new location without leaving a mark on the paper.

\begin{figure}
    \centering
    \includegraphics[width=0.26\linewidth]{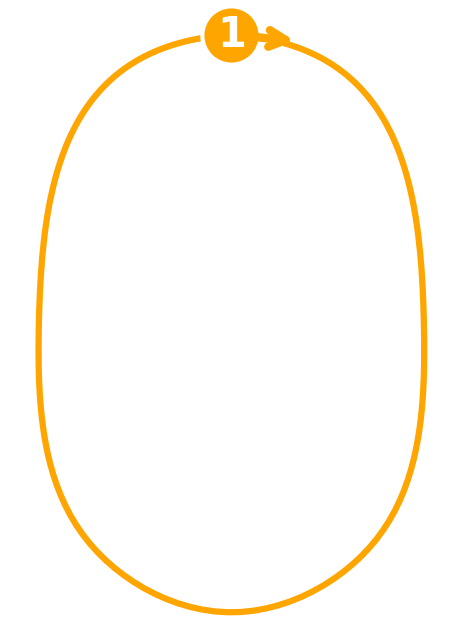}
    ~\includegraphics[width=0.23\linewidth]{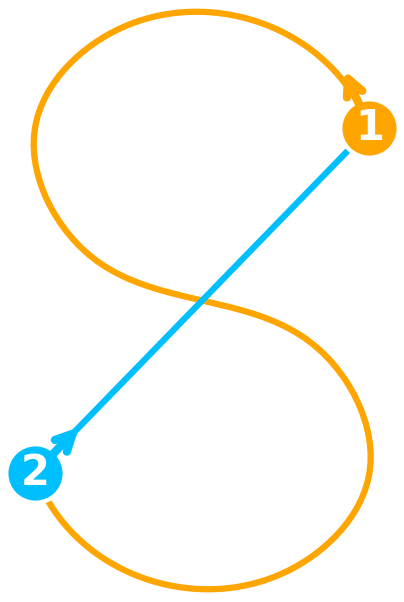}
    ~\includegraphics[width=0.23\linewidth]{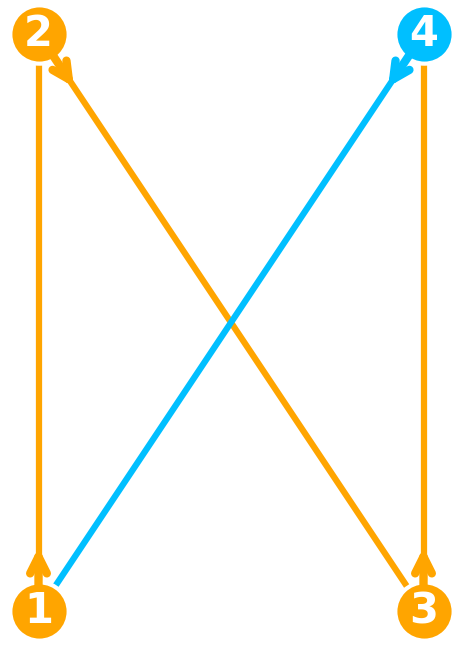}
    ~\includegraphics[width=0.23\linewidth]{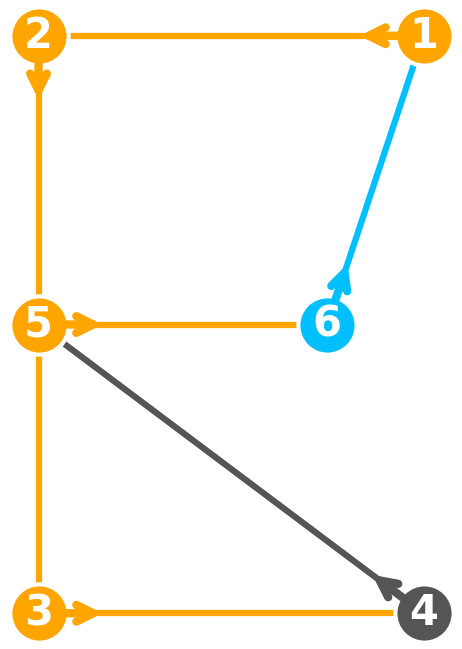}
    \caption{
    The writing path for the letters O, S, N, and E. Colors yellow, gray, and blue represent lit, dark, and return segments, respectively.
    }
    \label{fig:writing_path}
\end{figure}


As there are multiple ways to write a letter,
one may design different writing paths to draw a letter. To create a writing path, an author designs the lit segments by defining a sequence of control points that outline the desired segment form. 
MOPL processes the segments to create a complete writing path. 
First, it adds dark segments in between consecutive lit segments if the end point of the first segment is not aligned with the start point of the second segment.
The dark segments only include the two points, and they are only inserted between two lit segments. 
If the first and last lit segments are not connected, MOPL adds a final dark segment called the Return Segment to enable continuous animation. This segment connects the end of the last segment back to the beginning of the first, creating a closed loop that can be repeated seamlessly. 
Finally, MOPL smooths the segments to create a continuous writing path. It creates a straight line for two-point segments and uses B-splines to create a curve for segments with more than two control points.
Due to lack of space, we refer the interested reader to our github repository \url{https://github.com/flyinglightspeck/FLS} for the 
the number of segments and their type for all English letters. Figure \ref{fig:writing_path} shows the writing path for letters O, S, N, and E.



While a writing path is spatial, its \textit{Trajectory} defines how that spatial data is executed over time.
It is a time-sampled motion plan structured as a sequence of \textit{Fragments}. Each fragment corresponds to a segment in the writing path and contains a list of discrete \textit{Setpoints}.
A setpoint is an instruction that specifies the FLS's required position and velocity at a moment in time. 
The interval of time between two consecutive setpoints is fixed at $\tau$.

\begin{figure}
    \centering
    \includegraphics[width=0.29\linewidth]{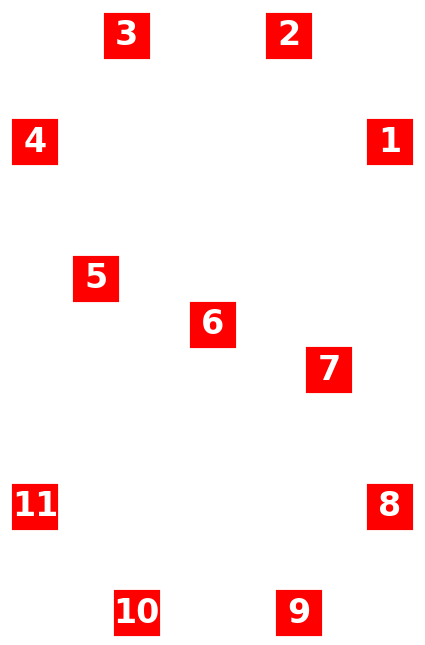}
    ~\includegraphics[width=0.31\linewidth]{figs/S_segments.png}
    ~\includegraphics[width=0.285\linewidth]{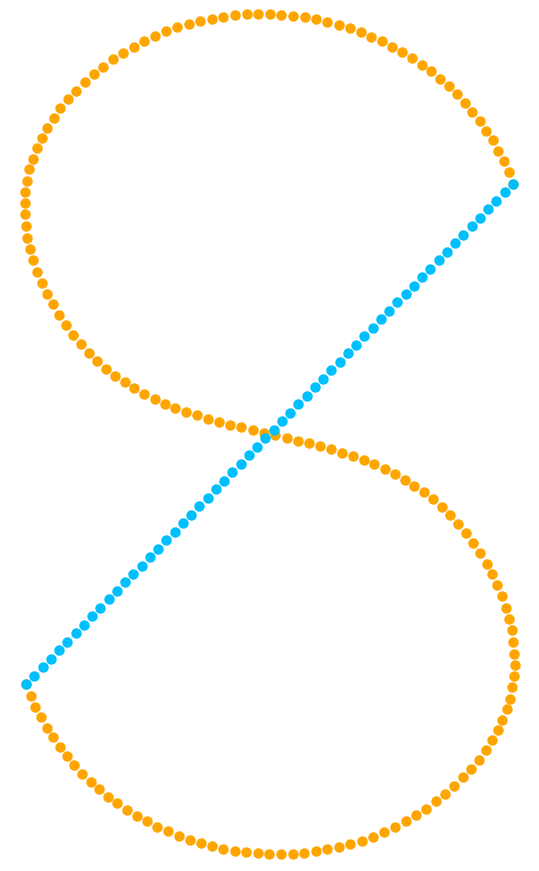}
    \caption{
    On the left, human authored discrete control points in red define the shape of the letter S. MOPL interpolates these points to form a smooth writing path consisting of one lit segment and one return segment in the middle. It computes a trajectory consisting of a sequence of set points separated by $\tau$ interval of time. 
    }
    \label{fig:trajectory}
\end{figure}

MOPL converts a writing path to a trajectory by transforming each segment of the writing path to a fragment in the trajectory. The fragment inherits the segment type, denoted as lit, dark, or return.
The transformation is governed by FLS's speed and acceleration.
It computes the duration
of a fragment based on the segment's length and the number of setpoints by dividing the fragment’s duration by $\tau$. It models the motion of FLS as it traverses the segment, including the three potential phases of motion: acceleration, moving at maximum speed, and deceleration to a stop. 
For example, when tracing the first lit segment of letter N (vertical line from bottom to top), the FLS accelerates from the start point, moves with maximum speed, and decelerates to stop at the end. To compute the setpoints, the MOPL determines the distance traveled at each time step. Then, it maps each distance to a point on the segment, producing the 3D position for each setpoint. Finally, it computes the setpoint's velocity as the change in position between consecutive setpoints divided by the $\tau$. The resulting list of setpoints constructs a fragment. The MOPL generates a fixed number of setpoints per unit of time.
However, the speed of the FLS is different at different times, e.g., when it accelerates from a full stop compared to when it travels at its maximum speed. 
This may result in a different number of setpoints per unit of length for different portions of a fragment.


Depending on the letter's shape, two strategies were used to create the trajectories.
The trajectories for letters E and N were generated with a max speed of $0.75 m/s$ and an acceleration of $0.1 m/s^2$. This ensures the FLS comes to a full stop before entering a sharp bend, preventing deviation from the trajectory.
For letters S and O, a constant speed of $0.25 m/s$ was used without acceleration and deceleration, since these letters do not have sharp bends.
Figure \ref{fig:trajectory} shows the conversion of control points to the writing path and then the trajectory for letter S.

An FLS follows the trajectory of a letter to illuminate it. During lit fragments, it turns its LED on, and keeps the LED off during dark fragments. The trajectory may also be rendered using an animation authoring tool such as Blender. We use both in Section~\ref{sec:eval}.


\section{Evaluation}\label{sec:eval}
We evaluated the FLS illumination of four letters both quantitatively and qualitatively.
The latter employs an IRB approved human subject study.  
We describe these in turn.
\subsection{Quality Metric}
\begin{table*}[htbp]
\centering
\caption{Trajectory Error, $\Delta_{Traj}$, in millimeter for FLS illuminations of four different letters.  Duration is in seconds.}
\label{tab:trajerr}
\begin{tabular}{c|cccc}
\hline
\textbf{Letter} & \textbf{Blender Animation} & \textbf{FLS Illumination} & \textbf{Illumination Duration} & \textbf{$\Delta_{Traj}$ (mm)}\\
\hline
S & \url{https://youtu.be/tdZ9qH9T3BA} & \url{https://youtube.com/shorts/qPW3V84tzOI} & 7 & 51\\
E & \url{https://youtu.be/kMSZh9U1qTE} & \url{https://youtube.com/shorts/G1pu7I2KoIM} & 25 & 42\\
O & \url{https://youtu.be/YfQDoi8uBdI} & \url{https://youtube.com/shorts/emOO9UJIrrw} & 8 & 56\\
N & \url{https://youtu.be/cgEES3tRKoc} & \url{https://youtube.com/shorts/yg7SNPl7fKo} & 18 & 46\\
\hline
\end{tabular}
\end{table*}

We define {\em Trajectory Error}, $\Delta_{Traj}$, to quantify the accuracy of the FLS illumination of a trajectory. 
It measures the deviation between the FLS's actual flight path and the intended trajectory of a letter.
A lower $\Delta_{Traj}$ signifies more accurate execution of the desired trajectory, with a value of 0 indicating a perfect match.
We capture the FLS's flight path as a sequence of 3D positions using the Vicon motion tracking system during FLS flight and compare it against the intended trajectory.
To compute the $\Delta_{Traj}$, we first normalize the FLS's flight path and the desired trajectory by aligning their centroids on the origin. Then, we calculate the shortest Euclidean distance to the trajectory for each point in the FLS's flight path. This is the error at each instance of time. Finally, we calculate the $\Delta_{Traj}$ as the Root Mean Square Error (RMSE) of all calculated errors:
\begin{equation}
\Delta_{Traj} 
= \sqrt{ \frac{1}{N} \sum_{i=1}^{N} \left\| \mathbf{x}_i - NN(x_i, P) \right\|^2 }
\end{equation}
Where $NN(\mathbf{x}_i, P) = argmin_{x'\in P}\left\|x_i-x'\right\|$ is the closest point on desired trajectory $P$ to $x_i$, $x_i$ is the 
position at time step $i$, and $N$ is the total number of sampled time steps 
in the FLS's flight path.
We used Vicon with 18 cameras to quantify $x_i$.
Table~\ref{tab:trajerr} shows the YouTube video link of the Blender animation of the intended trajectory, the YouTube video link of the FLS illumination, the FLS illumination duration of the letter, and its $\Delta_{Traj}$ for four letters.

\subsection{Human Subject Study}


\begin{table}[htbp]
\centering
\caption{Aggregate Metrics.  Time and duration in seconds.}
\label{tab:aggmetrics}
\begin{tabular}{c|c|c|c}
\hline
\multirow{2}{*}{\textbf{Letter}} & \textbf{Percentage} & \textbf{Average} & \textbf{Average} \\
 & \textbf{Detected} & \textbf{Confidence} & \textbf{Duration} \\
\hline
S & 20\%* & 4 & 102.9 \\
E & 30\% & 3 & 92.98 \\
O & 45\% & 3.7 & 93.97 \\
N & 65\%* & 4 & 105.68 \\ 
\hline
\multicolumn{4}{l}{\small * Significant difference, p=0.011.}
\end{tabular}
\end{table}

We evaluated the qualitative detection of the FLS illuminated English letters by conducting an IRB-approved human subject study with 20 participants:  
5 female, ages 25–30; 15 male, ages 25–32.
We illuminated 4 English letters:
O, N, S, and E. 
We chose these based on their number of segments, hypothesizing that a letter with more segments will be harder to detect by a human subject.
We recorded 3 repeated FLS illuminations of each letter in a dark room.
The time for one illumination of a letter (without the dark return segment at the end) is shown in the fourth column of Table~\ref{tab:aggmetrics}.

We used Latin Square~\cite{fisher1935design} to compute the following 4 sequences of letters that minimize the influence of the ordering effect:
NOSE, OSEN, SENO, and ENOS.
We assigned a sequence to a subject in a round robin manner. Five subjects to each sequence.
A subject is seated at the table, provided with a description of what is to happen, and asked to acknowledge a consent statement.
Next, they are presented with a divided screen.
On one side, the video of the letter loops continuously.
On the other side, the subject is asked "What did you see? If unsure, provide your best guess."
They type their answer in a text box.
The generic nature of the question was intentional as we did not want to bias the subjects\footnote{An alternative question that may have biased the subjects would have been:  What letter did you see being drawn?}.
Below the question, the subject rates their confidence in their reply, ranging from 1 (Not Confident at all) to 5 (Very Confident).
Subsequently, they are presented with the next letter in their assigned sequence on the divided screen.
This process repeats until the entire sequence is presented to the subject.
We time the subject from the start of the video until the subject clicks the button for the next letter.


Table~\ref{tab:aggmetrics} shows the percentage of subjects that detected each letter, the average confidence of those who detected the letter correctly, and their average duration of time to detect the letter.
It is interesting to note that the latter is independent of the illumination duration, see Table~\ref{tab:trajerr}.
For example, 
the FLS illumination of Letter E is 17 seconds longer than that of Letter O.  However, the average duration for a human subject to detect E is approximately the same as Letter O.
Letter S has the lowest and N has the highest detection.
Some users reported Letter S as Number 8.
If one qualifies these as having detected S, then the percentage detection of S increases to 55\%, 11 subjects. 
With the Letter O, we accepted the response "circular" as letter\footnote{If "circular" is not accepted, then the percentage detection drops to 10\%, 2 subjects.} O.
We were surprised\footnote{Originally, we hypothesized Letter O to have the highest percentage detection (with the circular response) because it consists of one segment and is simplest.} to see that Letter N has the highest detection rate.
Letter S with fewer total segments (2) than Letter E (6 segments) has a lower detection rate. However, its subject confidence is higher, 4 vs 3.

\begin{table}[htbp]
\centering
\caption{Sequences, detection rates, and duration in seconds.}
\label{tab:seq}
\begin{tabular}{c|c|c}
\hline
\textbf{Sequence} & \textbf{\% Detected} & \textbf{Average Duration} \\
\hline
ENOS & 50\% & 54.26* \\
NOSE & 45\% & 139.98* \\
SENO & 45\% & 106.59 \\
OSEN & 20\% & 106.84 \\
\hline
\multicolumn{3}{l}{\small * Significant difference, p=0.021.}
\end{tabular}
\end{table}

\begin{figure}
    \centering

    \includegraphics[width=0.5\linewidth]{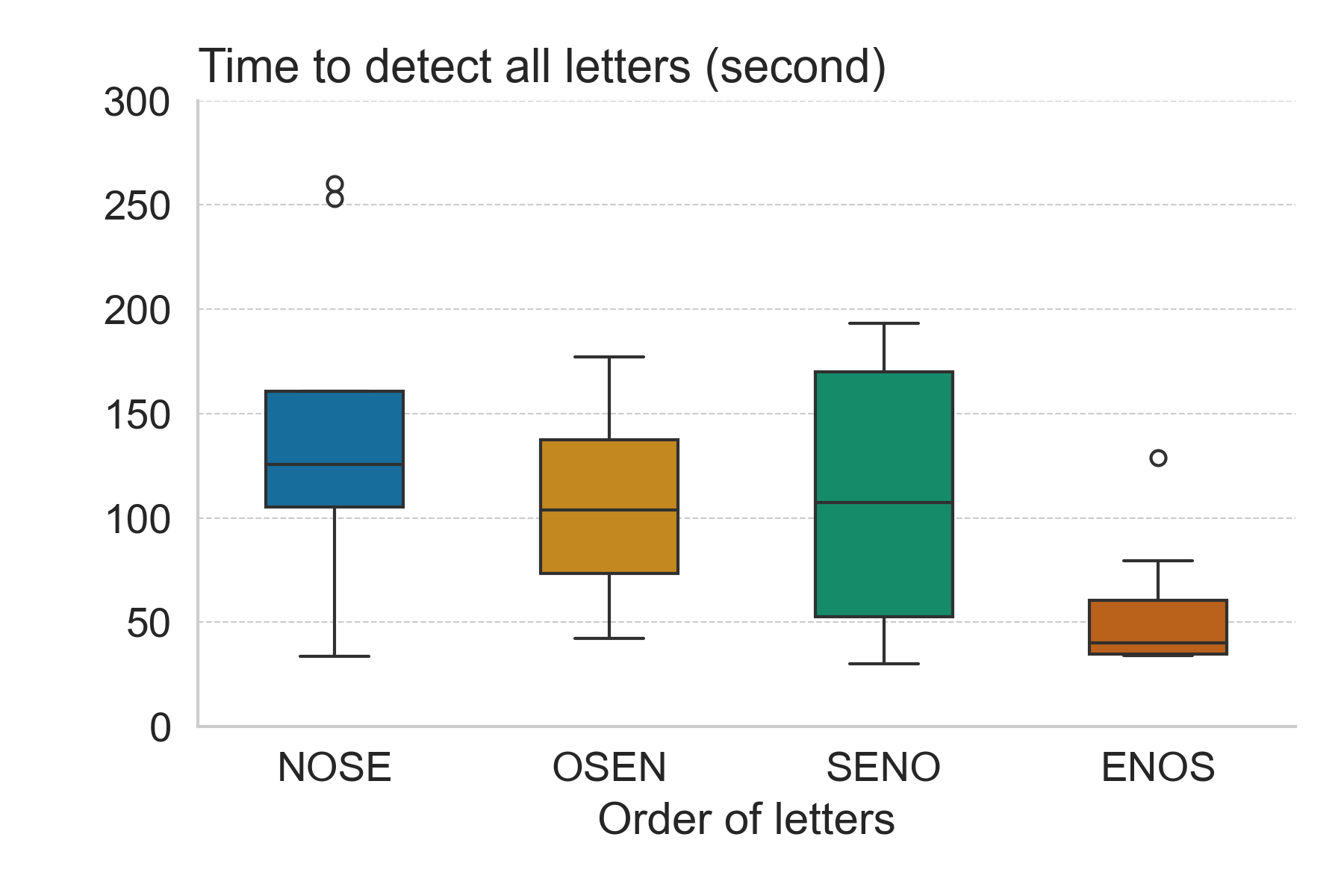}
    ~\includegraphics[width=0.5\linewidth]{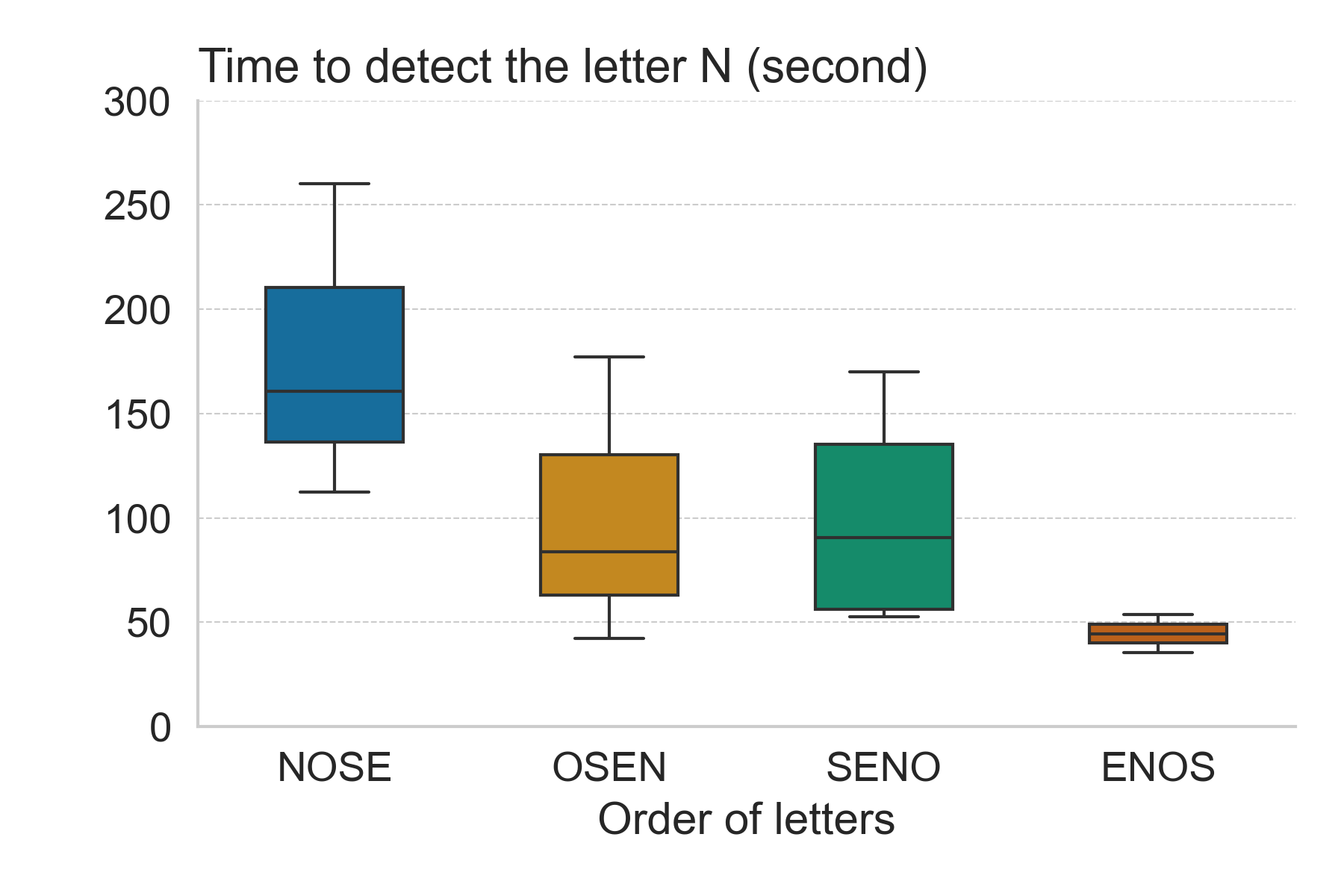}
    \caption{Elapsed time to detect all letters and N correctly.}\label{fig:detectionTime}
\end{figure}

The sequence of the FLS illuminated letter is important.
It impacts both the detection of letters by the subjects and how fast they detect a letter.
See Table~\ref{tab:seq}.
ENOS had the highest percentage of detected letters at 50\%.
The subjects were quickest in detecting letters with this sequence.
Almost 2x faster than the second pattern with the highest detection rate.
Figure~\ref{fig:detectionTime} shows the box plot for the time required by the subjects who detected all letters and Letter N for the different sequences.
They show (a) the slowest detection time for NOSE and (b) the lowest variation in detection time across the subject for the ENOS sequence.

\begin{table}[htbp]
\centering
\caption{Number of subjects detecting the correct letters.}
\label{tab:numsubjects}
\begin{tabular}{c|c|c}
\hline
\textbf{\# of Correct Letters} & \textbf{\# of Subjects} & \textbf{\% of Subjects} \\
\hline
0 (All Wrong) & 5 & 25\% \\
1 & 5 & 25\% \\
2 & 4 & 20\% \\
3 & 5 & 25\% \\
4 (All Correct) & 1 & 5\% \\
\hline
\end{tabular}
\end{table}

Table~\ref{tab:numsubjects} shows the number of subjects that detected 0 or more letters.
While one female subject detected all four letters with the NOSE pattern, a quarter of the subjects did not detect a letter, i.e., detected zero letters.
Among the latter, some subjects were puzzled by the task and its question.  

There was a significant relationship between the letter presented and its detection rate (chi-square test of independence: $X^2(3, 80)=9.58, p=0.023$). Specifically, the letter 'N' was detected significantly more often (65\%) than the letter 'S' (20\%) (Pairwise chi-square, p=0.011).
Additionally, the order in which letters were presented had a significant effect on detection duration (one-way ANOVA, $F(3,28)=3.34,p=0.034$). 
Post-hoc tests demonstrated that detection times were significantly shorter for the 'ENOS' pattern compared to the 'NOSE' pattern (Tukey's HSD, p=0.021).

\section{Related Work}\label{sec:related}
To the best of our knowledge, the use of an FLS to illuminate English letters is novel and has not been described in the literature.
The most relevant study is~\cite{dronepaint}.
It describes the use of hand gestures to control drones.
A user may use different sequences of hand gestures to cause a Crazyflie to take off, draw, erase, and land.
The user may use their index finger to draw an English letter in front of a camera.
DronePaint's hand tracking and Deep Neural Network (DNN) computes the trajectory and uses a Vicon system to instruct the Crazyflie to follow along the trajectory.
This study reports 56 millimeter error for the gesture-drawn trajectories. 
It does not report on the error introduced by the Crazyflie as the Vicon system is known to be highly accurate~\cite{preiss2017}.
Our work is different in several ways.  
First, there are no hand gestures.  
Second, the intended trajectories for the English letters are concise and drawn using Blender.
Third, we implement a decentralized localization technique using a camera on the FLS and IR markers on the floor.
DronePaint uses a centralized localization technique, the Vicon system.
Vicon requires a marker on the Crazyflie, IR cameras on the wall to track the marker, and a computer to process images of the marker from the cameras to compute trajectories that it transmits to the Crazyflie to follow.
Our FLS has its own local computing (Raspberry Pi 5) to process the images from its on-board camera to compute and control its trajectory.

\section{Future Research}\label{sec:future}
Our future research directions are as follows.
First, we want to minimize the trajectory error.
This error, 42-56 millimeter, is too high for certain letters, resulting in a very low human detection rate, e.g., 20\% for Letter S.
We plan to identify its source (e.g., Flight Controller, position estimator, the mechanical imperfections of the FLS) and address it to minimize the error.
Second, we are extending this study to use multiple FLSs to illuminate a letter.
A key consideration is what fragments of a trajectory the different FLSs should illuminate, and what is the ideal number of FLSs that results in the best detection rate.
Finally, for optimal illumination, an FLS must prioritize artistic considerations, carefully choosing light intensity and color to create an aesthetically pleasing experience for subjects.

\begin{acks}
This research was supported in part by the NSF grants CMMI-2425754 and IIS-2232382.
\end{acks}

\balance


\bibliographystyle{ACM-Reference-Format}
\bibliography{refs}

\end{document}